\documentstyle[11pt,aaspp4]{article}
\begin{document}
\newcommand{\msun}{M_{\odot}}
\newcommand{\lsun}{L_{\odot}}
\newcommand{\zsun}{Z_{\odot}}
\newcommand{\kms}{\, {\rm km\, s}^{-1}}
\newcommand{\cm}{\, {\rm cm}}
\newcommand{\gm}{\, {\rm g}}
\newcommand{\erg}{\, {\rm erg}}
\newcommand{\mpc}{\, {\rm Mpc}}
\newcommand{\seg}{\, {\rm s}}
\newcommand{\angs}{\, {\rm \AA}}
\newcommand{\hz}{\, {\rm Hz}}
\newcommand{\hi}{H\thinspace I\ }
\newcommand{\hii}{H\thinspace II\ }
\newcommand{\heii}{He\thinspace II\ }
\newcommand{\heiii}{He\thinspace III\ }
\newcommand{\nhi}{N_{HI}}
\newcommand{\lya}{Ly$\alpha$ }
\newcommand{\etal}{et al.\ }
\newcommand{\yr}{\, {\rm yr}}
\newcommand{\eq}{eq.\ }
\def\arcsec{''\hskip-3pt .}

\title{La Frontera a Alto Corrimiento al Rojo: Historia de la
Formaci\'on de las Galaxias}
\author{Jordi Miralda Escud\'e$^{1}$}
\affil{University of Pennsylvania, Dept. of Physics and Astronomy,
David Rittenhouse Lab.,
209 S. 33rd St., Philadelphia, PA 19104}
\authoremail{jordi@llull.physics.upenn.edu}
\affil{$^{1}$ Alfred P. Sloan Fellow}

\begin{abstract}

  This review presents a brief discussion of the theory of Cold Dark
Matter of structure formation in the universe, describing the main
processes determining the power spectrum, the non-linear gravitational
collapse, the formation of galaxies and the evolution of the
intergalactic medium. Recent advances in the observations of high
redshift galaxies, and their interpretation in the context of this
theory, are then summarized.

  Este art\'\i culo de revisi\'on presenta una breve exposici\'on de
la teor\'\i a de la Materia Invisible Fr\'\i a de la formaci\'on de
estructura en el universo, describiendo los procesos m\'as importantes
que determinan el espectro de potencia, el colapso gravitatorio
no lineal, la formaci\'on de galaxias y la evoluci\'on del medio
intergal\'actico. Resumimos tambi\'en avances recientes en las
observaciones de galaxias a alto corrimiento al rojo, y la
interpretaci\'on en el contexto de esa teor\'\i a.

\end{abstract}


\section{Introducci\'on}

  Ciertas cuestiones que han resultado en grandes avances en la historia
de la astronom\'\i a y la cosmolog\'\i a tienen, adem\'as, un inter\'es
especial para la humanidad porque tratan sobre fen\'omenos
totalmente imprescindibles para nuestra existencia en el universo: la
fuente de energ\'\i a del Sol, la formaci\'on de sistemas planetarios, y
la s\'\i ntesis de los elementos en el interior de las estrellas son
ejemplos que acuden a la mente. La cosmolog\'\i a moderna, despu\'es de
habernos encaminado por vez primera al estudio cient\'\i fico del origen
del universo observable en su totalidad, en el marco de la teor\'\i a de
la Gran Explosi\'on, trae consigo tambi\'en otra cuesti\'on dentro de esa
categor\'\i a: el origen de las fluctuaciones primordiales y la
formaci\'on de las galaxias. El universo debe empezar con condiciones
iniciales pr\'acticamente homog\'eneas para resultar en el estado
presente de homogeneidad a gran escala; pero la existencia de
fluctuaciones de densidad primordiales, de una amplitud $\delta\rho /
\rho \sim 10^{-5}$, es absolutamente necesaria para la formaci\'on de
las galaxias. La gravedad puede amplificar y llevar las fluctuaciones
iniciales al colapso no lineal, pero no puede crearlas. El proceso que
gener\'o esas fluctuaciones en el universo primitivo, sumido todav\'\i a
en el misterio, di\'o lugar a la gran diversidad del universo no lineal,
incluida la existencia de la vida, y evit\'o la continuaci\'on
indefinida del universo lineal y homog\'eneo, conteniendo \'unicamente
un mar de radiaci\'on y \'atomos de hidr\'ogeno y helio.

  Estamos actualmente en la era de exploraci\'on y descubrimiento hacia
la frontera de alto corrimiento al rojo o, en otras palabras, de las
mayores distancias des de las que es posible recibir mensajes en el
universo. Dada la velocidad de la luz, las grandes distancias
proporcionan a la cosmolog\'\i a la evidencia de la historia pasada de
la formaci\'on de las galaxias. Durante las \'ultimas tres d\'ecadas,
nuestra visi\'on del pasado del universo ha progresado continuamente con
el descubrimiento de galaxias y n\'ucleos activos a distancias cada vez
mayores, el estudio de fondos de radiaci\'on c\'osmica a distintas
frecuencias, y el an\'alisis de espectros de absorci\'on del hidr\'ogeno
intergal\'actico interpuesto en la direcci\'on de fuentes luminosas.
Presentamos en este art\'\i culo de revisi\'on un breve resumen del
estado actual de la teor\'\i a y observaciones sobre la formaci\'on de
las galaxias y la evoluci\'on del medio intergal\'actico. No podemos
hacer justicia en ese breve art\'\i culo al inmenso n\'umero de trabajos
publicados en diversos temas de gran impacto sobre el estudio de la
formaci\'on de las galaxias. Entre otros art\'\i culos de revisi\'on,
Ellis (1997) describe el estudio de galaxias d\'ebiles, y Rauch (1998)
presenta una excelente exposici\'on de observaciones y teor\'\i as del
bosque Ly$\alpha$ de hidr\'ogeno intergal\'actico. Cabe recomendar
asimismo varios art\'\i culos en libros de conferencias recientes de
gran utilidad para ponerse al d\'\i a en ese campo: v\'ease, por
ejemplo, Madau 1999, Steidel 1998a,b.

\section{Teor\'\i a de la Materia Invisible Fr\'\i a}

  De entre los distintos modelos propuestos de formaci\'on de galaxias,
la teor\'\i a de la Materia Invisible Fr\'\i a (que abreviaremos MIF; en
ingl\'es, Cold Dark Matter) ha resultado ser la de mayor \'exito, y
claramente favorecida por las observaciones. La teor\'\i a postula que
la materia invisible, cuya presencia se deduce de las determinaciones
din\'amicas de la masa de galaxias y c\'umulos (p.e., Trimble 1987),
consiste en
objetos o part\'\i culas ``fr\'\i as'', sin dispersi\'on de velocidades
inicial, y que las fluctuaciones primordiales de densidad son
adiab\'aticas (es decir, manteni\'endose constante la raz\'on de la
densidad de fotones, bariones y materia invisible), Gausianas, y con un
espectro de potencias invariante de escala (v\'ease por ejemplo,
Blumenthal \etal 1984, Ostriker 1993). Generalmente, se supone que las
fluctuaciones fueron generadas en un per\'\i odo de inflaci\'on por
alg\'un proceso que se mantuvo constante mientras el rango de escalas
observable en el presente cruzaba el horizonte de acontecimientos, lo
cual implica la invariancia de escala.

  Inicialmente, la teor\'\i a MIF se consider\'o de forma casi exclusiva
dentro del modelo cosmol\'ogico con densidad de materia cr\'\i tica,
$\Omega \equiv \rho/\rho_{crit} = 1$ (donde $\rho$ es la densidad media
de materia y $\rho_{crit}$ es la densidad cr\'\i tica). Desde los
inicios de la cosmolog\'\i a observacional, las medidas de la densidad
media del universo indicaron un valor $\Omega < 1$ (v\'ease, p.e. Gott
\etal 1974). Este resultado observacional se ha mantenido hasta la
actualidad: la densidad media de luz en la banda B es en el presente
$\rho_B \simeq 2\times 10^8\, h \lsun\mpc^{-3}$ (Zucca \etal 1997 y
referencias inclu\'\i das; utilizamos aqu\'\i$~ H_0 = 100 h
\kms\mpc^{-1}$), y la raz\'on masa-luminosidad en diversos sistemas
colapsados, desde galaxias individuales a los mayores c\'umulos, medida
a radios suficientemente grandes, tiende generalmente a valores $150 h
\lesssim M/L_B \lesssim 500 h$ (p.e., Bahcall, Lubin, \& Dorman 1995),
resultando en una densidad de materia $\bar\rho \simeq 5\times 10^{10}
h^2 \msun\mpc^{-3}$, y $\bar\rho/\rho_{crit} = \Omega \simeq 0.2 $.

  El valor de $\Omega$ as\'\i$~$ deducido a partir de la densidad
luminosa del universo puede verse afectado por el efecto del sesgo de
las galaxias.  En principio, la mayor parte de la masa del universo
podr\'\i a ubicarse en los grandes vac\'\i os, sin pertenecer a ninguna
estructura virializada conteniendo galaxias donde la masa total pueda
determinarse a trav\'es de los m\'etodos din\'amicos habituales. Esa fue
la posibilidad bajo la cual el modelo Einstein-de Sitter ($\Omega=1$)
fue forzado a resguardarse de la evidencia observacional. Sin embargo,
varios avances durante la \'ultima d\'ecada han confirmado un valor de
$\Omega \simeq 0.3$. En los c\'umulos de galaxias de mayor masa, la
fracci\'on de la masa constitu\'\i da por bariones deducida por la
intensidad y el espectro de rayos X emitidos por el gas caliente es
$f_{bar} \simeq 0.06 h^{-3/2}$ (S. D. M. White \etal 1993; D. A. White
\& Fabian 1994). Dada la densidad de bariones deducida de la
teor\'\i a de nucleos\'\i ntesis primordial, $\Omega_b \simeq 0.019
h^{-2}$ (p.e., Burles \& Tytler 1998), y el hecho de que la
fracci\'on de masa bari\'onica en los c\'umulos m\'as masivos debe ser
representativa de la fracci\'on media en el universo (White \etal 1993),
deducimos un valor $\Omega\simeq 0.3 h^{-1/2}$. Al mismo tiempo, las
observaciones recientes de curvas de luz de supernovas tipo Ia deducen
el mismo valor de $\Omega$, concluyendo adem\'as que la expansi\'on
del universo est\'a acelerando de la forma esperada en el modelo del
universo con geometr\'\i a espacial plana, donde la constante
cosmol\'ogica proporciona la densidad de energ\'\i a adicional
necesaria para llegar a la densidad cr\'\i tica (Perlmutter \etal 1998,
Riess \etal 1998). Otros m\'etodos de medir el valor de $\Omega$ son,
por lo general, consistentes con este resultado (por ejemplo, el valor
$\Omega \simeq 0.3$ con una constante cosmol\'ogica $\Lambda=1-\Omega$
tambi\'en es favorecido por el valor de la constante de Hubble y la edad
de las estrellas m\'as antiguas en c\'umulos globulares).
Varias observaciones
en el futuro pr\'oximo (tales como supernovas Tipo Ia y fluctuaciones en
la radiaci\'on c\'osmica de fondo) deber\'an permitir medir $\Omega$
con mayor precisi\'on, y clarificar si la reciente aceleraci\'on del
universo se debe a una constante cosmol\'ogica o al modelo m\'as general
de un campo escalar con presi\'on negativa (p.e., Peebles \& Vilenkin
1998 y referencias inclu\'\i das).

  Una vez adoptamos el modelo cosmol\'ogico deducido a partir de esas
observaciones, las predicciones de la ter\'\i a MIF se ajustan bien a
los datos observacionales sobre estructura a gran escala, tales como la
funci\'on de correlaci\'on espacial de las galaxias, la abundancia
de c\'umulos de galaxias, la evoluci\'on en CR de esas cantidades, y
las fluctuaciones en la radiaci\'on c\'osmica de fondo recientemente
detectadas. Para obtener esas predicciones, es preciso calcular el
espectro de potencias de las fluctuaciones primordiales. 

  El espectro de potencias de las fluctuaciones de densidad se obtiene
en la teor\'\i a MIF calculando la evoluci\'on en el r\'egimen lineal
de esas fluctuaciones una vez cruzan el horizonte (para art\'\i culos
de revisi\'on, v\'ease Efstathiou 1990, Bond 1996, Bertschinger 1996).
En el l\'\i mite de
peque\~nas escalas, las fluctuaciones entraron en el horizonte durante
la \'epoca en que la densidad media del universo era dominada por
radiaci\'on. Los bariones est\'an entonces ligados a la radiaci\'on, y
la presi\'on radiativa resulta en una oscilaci\'on de las fluctuaciones,
impidiendo el crecimiento gravitatorio. La materia invisible no
interact\'ua con la radiaci\'on, y por lo tanto aumenta sus
fluctuaciones de densidad. No obstante, ese crecimiento gravitatorio es
muy lento incluso en la ausencia de acoplamiento con la radiaci\'on,
cuando la radiaci\'on domina la densidad de energ\'\i a.

  Dado que la densidad de radiaci\'on en el presente es $\Omega_{rad} =
a T_{rad}^4/(\rho_{crit} c^2) = 2.47\times 10^{-5} h^{-2}$ (donde
$T_{rad}=2.76$ K es la temperatura de la radiaci\'on de fondo), la
\'epoca de igualaci\'on de las densidades de materia y radiaci\'on
corresponde al CR $1+z= 4.04\times 10^4\, \Omega h^2$. La longitud
del horizonte en esta \'epoca era del orden $c\, H^{-1}(z) = c H_0^{-1}
\Omega^{-1/2}\, (1+z)^{-3/2}$, y por lo tanto la longitud com\'ovil
era $L_{ig} = c H_0^{-1} \Omega^{-1/2} (1+z_{ig})^{-1/2} \simeq
15 (\Omega h)^{-1} h^{-1} \mpc$ (donde utilizamos la notaci\'on
habitual para la constante de Hubble en el presente, $H_0 = 100\, h
\kms\mpc^{-1}$, y $H(z)$ es la constante de Hubble a CR $z$).
A escalas com\'oviles mucho menores
que el horizonte en la \'epoca de igualaci\'on, la amplitud de
fluctuaciones de densidad debe conservar la forma de las fluctuaciones
primordiales cuando emergen del horizonte, ya que las
fluctuaciones empezaron a crecer solamente a partir de la \'epoca de
igualaci\'on, y el crecimiento en la \'epoca anterior fue muy lento.
Eso implica que la amplitud de fluctuaciones tiende a una constante
para el caso m\'as habitual de invariancia de escala, con $n=1$ (donde
el espectro primordial es $P(k) \propto k^n$). Pero en el
l\'\i mite de escalas mucho mayores, $L\gg L_{ig}$, las fluctuaciones
crecen en el r\'egimen lineal proporcionalmente al factor de escala,
$a \propto (1+z)^{-1}$, desde el momento en que entran en el horizonte;
dado que la longitud com\'ovil del horizonte es proporcional a
$(1+z)^{-1/2}$, el factor total de crecimiento a una \'epoca fija
debe ser proporcional a $L^{-2}$.

  Mostramos en la Figura 1 la dispersi\'on en la fluctuaci\'on
relativa de la masa total contenida dentro de una esfera de radio
com\'ovil $R$, cuando la evoluci\'on lineal de fluctuaciones
primordiales se extrapola hasta el presente. Conviene recordar
aqu\'\i$~$ que la distribuci\'on de probabilidad de esta masa es
Gausiana (dada la suposici\'on de que el campo de densidades es
Gausiano), con la dispersi\'on mostrada en la Figura 1. La l\'\i nea
s\'olida es para el modelo $\Omega = 0.3$, $\Lambda = 0.7$, $n=1$,
$H_0 = 65 \kms\mpc^{-1}$, y hemos normalizado las fluctuaciones a
$\sigma_8 \equiv < \sigma^2 >^{1/2} (R=8 h^{-1} \mpc) = 0.9$. La
l\'\i nea de rayas es para el modelo $\Omega=1$, $\Lambda = 0$, $H_0 =
50 \kms\mpc^{-1}$, y $\sigma_8 = 0.55$. Las curvas han sido calculadas
con las f\'ormulas presentadas por Hu \& Sugiyama (1996). Adem\'as del
comportamiento esperado en el l\'\i mite de peque\~nas y grandes
escalas, vemos tambi\'en que el modelo de la constante cosmol\'ogica
dispone de mayor potencia a grandes escalas, debido al mayor valor de la
longitud caracter\'\i stica $L_{ig} \propto (\Omega h)^{-1}$.

  La normalizaci\'on del espectro de fluctuaciones, dada por el
par\'ametro $\sigma_8$, debe considerarse como un par\'ametro ajustable
de la teor\'\i a. Su valor ha sido escogido aqu\'\i$~$ para reproducir
aproximadamente la abundancia de c\'umulos de galaxias en el presente
(Eke, Cole, \& Frenk 1996). Desde que las fluctuaciones de intensidad de
la radiaci\'on de fondo de microondas fueron detectadas por el
sat\'elite COBE, la normalizaci\'on viene fijada independientemente por
esas fluctuaciones (Bennett \etal 1996; Bunn \& White 1997). Esas dos
medidas de la amplitud concuerdan para el modelo con constante
cosmol\'ogica con $\Omega\simeq 0.3$, mientras que son claramente
conflictivas si se supone $\Omega=1$. Es notable el hecho de que cuando
se toma el valor de $\Omega$ deducido de las observaciones de la
geometr\'\i a global del universo y la densidad promedio de materia, la
teor\'\i a MIF predice correctamente la relaci\'on entre las
fluctuaciones de la radiaci\'on de fondo y la abundancia de c\'umulos
de galaxias, y concuerda adem\'as con la correlaci\'on espacial de
galaxias (Jing, Mo, \& B\"orner 1998 y referencias inclu\'\i das),
as\'\i$~$ como la evoluci\'on de la abundancia de c\'umulos en CR
(Bahcall \& Fan 1998).

\section{Colapso gravitatorio no-lineal: formaci\'on de halos}

  En la teor\'\i a MIF, la materia invisible empieza a formar objetos
virializados (o ``halos'') por colapso gravitatorio a partir de las
escalas m\'as peque\~nas, cuando la fluctuaci\'on de densidad alcanza
un valor $\sim 1$, entrando en el r\'egimen no lineal. Luego, cuando
las fluctuaciones a escalas mayores llegan al colapso, halos de mayor
masa se forman por mersi\'on de halos de masa menor colapsados
anteriormente. S\'olo podemos presentar aqu\'\i$~$ un resumen muy breve
de la teor\'\i a de formaci\'on de halos y galaxias; para una
discusi\'on mucho m\'as extensa, v\'ease White (1996).

  Un modelo anal\'\i tico de gran utilidad para entender el proceso del
colapso no lineal, llevando a la formaci\'on y mersi\'on de halos, es el
modelo de Press-Schechter (Press \& Schechter 1974).
El modelo se basa en la soluci\'on anal\'\i tica del colapso de una
perturbaci\'on de densidad constante con simetr\'\i a esf\'erica, en
cuyo caso el colapso al punto central sucede cuando se alcanza el valor
de la sobredensidad extrapolada linealmente
$\delta = \delta_c = 3/5 (3\pi/2)^{2/3} = 1.686$ (Peebles 1980). La
distribuci\'on de
la densidad promediada sobre una esfera de radio $R$ es una Gausiana
con dispersi\'on $\sigma(R)$ (Figura 1), y el modelo Press-Schechter
consiste en suponer que la fracci\'on de la masa del universo que ha
colapsado en halos de masa mayor que $M=(4\pi/3) \bar\rho R^3$ es
igual a ${\rm erf}\{\delta_c/[2^{1/2}\sigma(R)] \}$. N\'otese que esa
fracci\'on es en realidad dos veces la fracci\'on del volumen con
sobredensidad superior a $\delta_c$ en las condiciones iniciales.
Aunque la introducci\'on del factor 2 puede
justificarse mejor con un tratamiento en el espacio de Fourier
(Bond \etal 1991), el modelo de Press-Schechter debe considerarse
\'unicamente como una aproximaci\'on que se utiliza
frecuentemente debido a su gran simplicidad, y especialmente al hecho
de que sus predicciones se ajustan a resultados num\'ericos de la
abundancia de halos con bastante precisi\'on (p.e., Governato \etal
1998; a la pr\'actica, el par\'ametro $\delta_c$ puede considerarse
ajustable a los resultados num\'ericos, en vez de tomar el valor
requerido para el colapso esf\'erico).

  Una vez obtenida la abundancia de halos de masa mayor que $M$, se
obtiene por diferenciaci\'on la abundancia diferencial, y asignamos
a cada escala $R$ una dispersi\'on de velocidades para el halo
obtenida tambi\'en por el modelo esf\'erico: $\sigma^2= (3\pi/2)^{2/3}\,
[H(z_c) R]^2/2$, donde $H(z_c)$ es la constante de Hubble en el momento
del colapso. B\'asicamente, esa relaci\'on nos dice
que la dispersi\'on de velocidad de un c\'umulo es del orden de la
velocidad de expansi\'on de Hubble sobre la regi\'on com\'ovil desde
la cual el c\'umulo ha colapsado, en el momento del colapso. La
dispersi\'on de velocidad de un halo (o, an\'alogamente, el cociente de
la masa sobre el radio) es la propiedad de un halo que determina las
cantidades observables en galaxias y c\'umulos, que son la dispersi\'on
de velocidad de las galaxias, la temperatura del gas difuso en
equilibrio din\'amico en el halo (determinada con espectros de rayos X),
o la deflecci\'on gravitatoria de luz de fondo (medida en lentes
gravitatorias).

  En las Figuras 2(a,b), las tres l\'\i neas s\'olidas gruesas nos dan
la dispersi\'on de velocidades de un halo formado a partir de una
fluctuaci\'on de densidad inicial igual a $(1,2,3)-\sigma$, en funci\'on
del CR, para los mismos dos modelos utilizados en la Figura 1 (para
el modelo con constante cosmol\'ogica en la Fig. 2b, los valores de
$\delta_c$ y de la dispersi\'on de velocidades correspondiente a cada
escala $R$ deben ser modificados debido al efecto de repulsi\'on de la
constante cosmol\'ogica, y el ritmo de crecimiento lineal de
fluctuaciones es tambi\'en distinto; v\'ease p.e., Viana \& Liddle 1996).
De acuerdo con la distribuci\'on Gausiana, la fracci\'on de masa que ha
colapsado en halos de mayor dispersi\'on de velocidades es,
respectivamente, (64\%,10\%,0.6\%). Tomemos, por ejemplo, el momento
presente ($z=0$) para el modelo $\Omega=0.3$ (Figura 2b). Ese modelo
nos predice que un $0.6\%$ de la masa en el presente est\'a en
objetos colapsados con dispersi\'on de velocidad mayor que $1000 \kms$.
Esos halos se corresponden evidentemente con los c\'umulos de galaxias
m\'as masivos, y se formaron a partir del colapso de raras
fluctuaciones a escalas grandes, $R\simeq 15\, h^{-1}\mpc$, donde
$3\sigma(R)= \delta_c$. Las l\'\i neas de rayas nos dan el valor de la
masa de los halos (con un factor 10 en masa separando l\'\i neas
sucesivas), e indican que la masa de uno de esos c\'umulos es
$\sim 2\times 10^{15} \msun$. La mayor parte de la masa del universo
debe encontrarse en objetos formados a partir de fluctuaciones m\'as
habituales: una fluctuaci\'on $1-\sigma$ produce un halo con
dispersi\'on de velocidades $\sim 200 \kms$, un valor t\'\i pico de los
peque\~nos grupos de galaxias, tales como el grupo Local, donde residen
la mayor\'\i a de las galaxias en el universo. La masa indicada en la
Figura 2 se refiere siempre a la masa total del halo en el momento de
su formaci\'on. Por ejemplo, la galaxia de la V\'\i a L\'actea pudo
formarse a $z\simeq 2$, en un halo con $\sigma \simeq 150 \kms$ de masa
$M\simeq 10^{12}\msun$, y en el presente el Grupo Local est\'a
colapsando en un halo mayor. Los dos modelos en la Figura 2 han sido
normalizados para ajustar sus predicciones a la abundancia observada de
c\'umulos. 

  La dispersi\'on de velocidades puede relacionarse con la temperatura
del gas difuso en el halo cuando est\'a en equilibrio hidorst\'atico:
$kT/\mu = \sigma^2$. Mostramos tambi\'en la temperatura en el eje
vertical. Hemos utilizado aqu\'\i$~$ la masa media para el caso de gas
primordial ionizado: $\mu \simeq 0.6 m_H \simeq 10^{-24}\, {\rm g}$. 

  Una propiedad general de la teor\'\i a MIF es que, a medida que las
fluctuaciones colapsan a escalas cada vez mayores, los halos deben
mergerse y formar nuevos halos con mayor dispersi\'on de velocidad;
adem\'as, el ritmo de crecimiento de la masa y la dispersi\'on de
velocidad es mucho m\'as r\'apido a alto CR que en el presente. Eso
es una consecuencia directa de la forma del espectro de potencias en la
Figura 1: a peque\~nas escalas, la amplitud de fluctuaciones es
pr\'acticamente constante, y por lo tanto la masa de los halos aumenta
r\'apidamente cuando esas fluctuaciones empiezan a colapsar. Pero a
mayores escalas, la amplitud disminuye con mucha m\'as rapidez con la
escala, y por lo tanto las fluctuaciones a escalas grandes colapsan
mucho m\'as tarde.

\section{Formaci\'on de galaxias}

  Una pregunta central en la teor\'\i a de formaci\'on de galaxias ha
sido la siguiente: cu\'al es la causa de la diferencia entre una galaxia
y un c\'umulo de galaxias? Evidentemente, los c\'umulos de galaxias han
colapsado gravitatoriamente, pero deducimos a partir de observaciones de
rayos X que la mayor parte de la materia bari\'onica est\'a en forma de
gas difuso y caliente distribu\'\i do en el halo. La masa total del gas
difuso en un c\'umulo masivo puede superar las $10^{14} \msun$, pero no
se forman nunca galaxias de masa tan grande.

  Esa cuesti\'on fue investigada por Binney (1977), Rees \& Ostriker
(1977), Silk (1977), y especialmente en el contexto de la teor\'\i a
MIF por White \& Rees (1978). En el colapso gravitatorio de un halo, la
materia bari\'onica se calienta inicialmente en ondas de choque, hasta
la temperatura necesaria para permanecer en equilibrio hidrost\'atico.
Luego, para poder condensar a una galaxia central, es preciso que el gas
pierda su energ\'\i a t\'ermica por procesos de radiaci\'on. Si el
tiempo necesario para que toda la energ\'\i a t\'ermica de los bariones
sea radiada es corto comparado con el tiempo entre mersiones sucesivas
de halos, entonces el gas va a poder concentrarse hacia el centro del
halo, y el proceso de colapso gravitatorio puede proseguir hasta llegar
a la formaci\'on de estrellas (despues de fragmentaci\'on de nubes de
gas, y posiblemente de haber formado un disco debido a la conservaci\'on
del momento angular; v\'ease Fall \& Efstathiou 1980, Fall \& Rees 1985)
o de un objeto masivo central, dando lugar a un n\'ucleo activo. En
cambio, si el tiempo de enfriamiento es demasiado largo, nuevas
mersiones para formar halos de mayor masa van a calentar de nuevo el
gas, manteni\'endolo en estado difuso.

  La condici\'on que debemos analizar es, por consiguiente, que el
tiempo de enfriamiento del gas sea igual a la edad del universo (que es
aproximadamente el tiempo entre mersiones sucesivas). Los procesos de
enfriamiento m\'as importantes son la radiaci\'on bremsstrahlung, y la
excitaci\'on y ionizaci\'on colisional de iones. A temperaturas $T
\lesssim 10^5 K$, el enfriamiento por hidr\'ogeno y helio domina, pero
en el rango $10^5 K < T < 10^7 K$ los iones de elementos pesados son
importantes (p.e., Gaetz \& Salpeter 1983). En las figuras 2(a,b), las
l\'\i neas de puntos indican la dispersi\'on de velocidad de halos en
los que el tiempo de enfriamiento es igual a la edad del universo, para
tres valores de la metalicidad del gas, $(0, 0.1, 1) \zsun$. Por encima
de las l\'\i neas de puntos, el tiempo de enfriamiento es demasiado
largo y se predice que el gas caliente se mantiene en el halo, mientras
que por debajo de las curvas el gas puede enfriarse y formar una
galaxia. El ritmo de enfriamiento por radiaci\'on se ha calculado
suponiendo que una fracci\'on $\Omega_b/\Omega$ de la masa del halo
est\'a en forma de gas difuso, donde $\Omega_b = 0.019 h^{-2}$, de
acuerdo con las mediciones de la abundancia de deuterio (Burles \&
Tytler 1998), y suponiendo el valor de la sobredensidad media de los
halos en el modelo del colapso esf\'erico (igual a $18\pi^2$ para
$\Omega=1$). Si las galaxias pueden formarse en halos por debajo de las
l\'\i neas de puntos, obtenemos la predicci\'on de que hasta $z \gtrsim
4$, el colapso de un halo debi\'o resultar siempre en la formaci\'on de
una galaxia central, pero a menores CR los halos m\'as masivos pueden
contener solamente galaxias formadas anteriormente, y la mayor parte
del gas difuso no puede enfriarse. Mencionamos aqu\'\i$~$ que, incluso
en los c\'umulos de mayor masa, una parte del gas en la regi\'on
central (con mayor densidad) puede radiar m\'as r\'apidamente; la
raz\'on por la cual eso no da resultado a grandes tasas de formaci\'on
estelar en las galaxias centrales de c\'umulos masivos es una cuesti\'on
que permanece sin resolverse (Fabian, Nulsen, \& Canizares 1994).

  Vemos, pues, que la f\'\i sica de enfriamiento del gas proporciona una
explicaci\'on satisfactoria del l\'\i mite superior de la masa de las
galaxias. Otra cuesti\'on distinta es el l\'\i mite inferior, y eso nos
lleva a una transici\'on fundamental en el universo para la historia de
formaci\'on de las galaxias: la reionizaci\'on del medio
intergal\'actico.

\section{La reionizaci\'on del universo}

  En la teor\'\i a de la Gran Explosi\'on, la materia del universo debe
formar \'atomos por vez primera cuando la temperatura de la radiaci\'on
de fondo disminuye hasta $T\simeq 3000$ K, a $z\simeq 10^3$. La materia
intergal\'actica permanece posteriormente en estado at\'omico, hasta el
momento en que el colapso no-lineal de las perturbaciones lleva a la
formaci\'on de objetos que emiten radiaci\'on ionizante, y pueden
ionizar de nuevo el gas difuso.

  Las observaciones del espectro de fuentes luminosas a alto CR
(generalmente, cu\'asares) demuestran que el medio intergal\'actico fue
ionizado anteriormente a $z\simeq 5$. La luz de una fuente a longitudes
de onda menores que la l\'\i nea \lya de hidr\'ogeno at\'omico puede ser
absorbida por un \'atomo a lo largo de su trayectoria a trav\'es del
universo, en el punto en que la longitud de onda se ha corrido hasta
coincidir con la l\'\i nea \lya. Si una parte importante de la densidad
bari\'onica media del universo estuviera distribu\'\i da por el espacio
en forma at\'omica, la profundidad \'optica de absorci\'on ser\'\i a
enorme ($\sim 10^5$ a $z=3$), con lo cual el flujo deber\'\i a disminuir
a cero abruptamente a la longitud de onda de \lya (Gunn \& Peterson
1965). Lo que se observa en realidad es la presencia de m\'ultiples
l\'\i neas de absorci\'on que producen un decremento neto del flujo de
solamente un $\sim 30\%$ a $z=3$. Eso implica que el medio
intergal\'actico est\'a altamente ionizado, con una fracci\'on neutra de
$\sim 10^{-5}$. Esa diminuta fracci\'on del hidr\'ogeno intergal\'actico
puede explicar satisfactoriamente las propiedades de las l\'\i neas de
absorci\'on observadas (denominadas usualmente como el ``bosque \lya"),
originadas en las variaciones de la densidad del gas que no ha colapsado
todav\'\i a en halos virializados de gran densidad. Varios trabajos
recientes, utilizando modelos anal\'\i ticos y simulaciones num\'ericas,
han mostrado que la teor\'\i a MIF predice de forma gen\'erica que el
medio intergal\'actico ionizado debe dar lugar a ese bosque de
l\'\i neas (v\'ease McGill 1990; Bi, B\"orner, \& Chu 1992; Bi 1993;
Miralda-Escud\'e \& Rees 1993; Cen \etal 1994; Zhang \etal 1995, 1998;
Hernquist \etal 1996; Miralda-Escud\'e \etal 1996). El hecho de que la
densidad del gas neutro es proporcional al cuadrado de la densidad del
gas cuando se establece el equilibrio entre recombinaci\'on y
fotoionizaci\'on por el fondo c\'osmico ionizante es la causa de que
un medio continuo da lugar a un espectro de absorci\'on con la
apariencia de ``l\'\i neas'' individuales, cada vez que se encuentra un
m\'aximo de la densidad del gas a lo largo de la l\'\i nea de visual.
El espectro de absorci\'on tiene una escala de suavizamiento natural
de $\sim 20 \kms$, debido a la dispersi\'on de velocidad t\'ermica del
gas fotoionizado (con temperatura $T\simeq 2\times 10^4$ K). El tama\~no
transversal de las estructuras de absorci\'on \lya debe ser del orden
de su dispersi\'on de velocidad multiplicada por la edad del universo,
ya que tales estructuras no han tenido todav\'\i a tiempo de colapsar y
de llegar a un equilibrio hidrost\'atico; observaciones recientes del
tama\~no transversal en pares de cu\'asares (Bechtold \etal 1994;
Dinshaw \etal 1994) confirman esa predicci\'on.

  Qu\'e caus\'o la reionizaci\'on del universo? Dos mecanismos pueden
ionizar el gas intergal\'actico: fotoionizaci\'on, o ionizaci\'on
colisional una vez el gas ha sido calentado por ondas de choque
provenientes de alguna explosi\'on de gran energ\'\i a. La
fotoionizaci\'on es el m\'etodo m\'as eficiente para ionizar el medio de
baja densidad, ya que requiere menos energ\'\i a, y la radiaci\'on se
transporta con gran eficacia a todas las regiones del espacio. Los dos
tipos de fuentes de radiaci\'on ionizante conocidas que pueden formarse
tan pronto como los primeros halos colapsan a peque\~nas escalas son
estrellas y n\'ucleos activos (producidos por agujeros negros masivos en
el centro de una galaxia). La cantidad de estrellas necesaria para
ionizar el universo entero es solamente una parte muy peque\~na de todas
las estrellas que se han formado hasta el presente, y puede relacionarse
f\'acilmente con la metalicidad media producida. Ya que tanto los
elementos pesados provenientes de explosiones supernova como la
radiaci\'on ionizante son producidas por estrellas masivas, la raz\'on
de esas dos cantidades est\'a relativamente fijada:
caracter\'\i sticamente, una estrella de $30 \msun$ fusiona unas
$4\msun$ de hidr\'ogeno en la secuencia principal, obteniendo una
energ\'\i a $\sim 0.03 \msun c^2$, de la cual $\sim 0.01 \msun c^2$ se
emite en fotones ionizantes. Esa misma estrella produce unas $5\msun$ de
elementos pesados en la explosi\'on supernova. Si los fotones son
absorbidos por hidr\'ogeno intergal\'actico (necesitando una energ\'\i a
de $\sim 20$ eV para cada ionizaci\'on, o una fracci\'on $2\times
10^{-8}$ de la masa-energ\'\i a en reposo), la masa ionizada es de $\sim
5\times 10^5 \msun$, y por lo tanto la metalicidad media aumenta s\'olo
hasta $10^{-5}$ una vez se ha emitido un fot\'on ionizante para cada
bari\'on. Dado que la metalicidad media en el universo presente es mucho
mayor, es evidente que la primera generaci\'on de estrellas puede
f\'acilmente ionizar todo el universo (p.e., Couchman \& Rees 1986).

  Evidentemente, los cu\'asares pueden ser tambi\'en las fuentes
dominantes para la reionizaci\'on, dado que la eficiencia en convertir
la masa acretada por un agujero negro en radiaci\'on ionizante es
generalmente mucho mayor que para estrellas. Las observaciones de la
abundancia de cu\'asares y de la intensidad del fondo c\'osmico de
radiaci\'on ionizante a $z \lesssim 3$ indican que los cu\'asares son
probablemente los mayores contribuyentes de esa radiaci\'on c\'osmica,
pero la naturaleza de las fuentes causantes de la reionizaci\'on a mayor
CR es todav\'\i a incierta (Miralda-Escud\'e \& Ostriker 1990;
Madau 1991, 1992, 1999; Haardt \& Madau 1996; Rauch \etal 1997;
Haiman \& Loeb 1997, 1998; Miralda-Escud\'e, Haehnelt, \& Rees 1999).

\section{Las primeras galaxias}

  Las primeras galaxias donde pudieron formarse las primeras estrellas
y cu\'asares surgieron, en la teor\'\i a MIF, en los primeros halos
que colapsaron donde el gas pudo enfriarse. El hidr\'ogeno at\'omico
solamente empieza a radiar en l\'\i neas de excitaci\'on cuando la
temperatura supera los $10^4$ K; a temperaturas inferiores, las
colisiones con electrones t\'ermicos no tienen nunca la energ\'\i a
necesaria para excitaci\'on a los niveles at\'omicos $n=2$. El
hidr\'ogeno molecular proporciona la \'unica fuente de enfriamiento
a menores temperaturas. En la materia primordial, en ausencia total
de elementos pesados, el hidr\'ogeno molecular se forma a partir del
i\'on $H^{-}$ (Saslaw \& Zipoy 1967), el cual resulta a su vez de
colisiones de hidr\'ogeno con los electrones del
residuo de ionizaci\'on que permanece despu\'es de la \'epoca de
recombinaci\'on (Peebles 1968). Ese proceso de formaci\'on de
mol\'eculas es muy ineficaz (en el presente, el hidr\'ogeno molecular se
forma en la superficie de granos de polvo interestelar), y s\'olo
una peque\~na fracci\'on del hidr\'ogeno forma mol\'eculas.
El enfriamiento resultante s\'olo es suficiente para resultar en la
formaci\'on de galaxias cuando la temperatura de un halo supera los
$\sim 2000 K$ (Tegmark \etal 1997; Abel \etal 1998).

  En las Figuras 2(a,b), la l\'\i nea s\'olida a mayor CR indica la
m\'\i nima temperatura para que un halo formado a CR $z$ pueda enfriar
su gas y formar una galaxia, que hemos reproducido de Tegmark \etal
(1997). Vemos que la teor\'\i a MIF, con los par\'ametros obtenidos de
observaciones mencionadas antes, predice que las primeras estrellas
pudieron formarse a $z\simeq 20$, en halos de dispersi\'on de
velocidades de $\sim 5 \kms$ y masa total $\sim 10^7 \msun$. Esa
primera generaci\'on de galaxias, formadas a trav\'es del enfriamiento
por hidr\'ogeno molecular, form\'o probablemente una cantidad muy
peque\~na de estrellas debido a la ineficacia de esta forma de
enfriamiento (v\'ease tambi\'en Haiman, Rees, \& Loeb 1997). En esta \'epoca,
el ritmo de mersiones es muy elevado y la masa y dispersi\'on de
velocidad de los halos aumenta r\'apidamente con el tiempo;
as\'\i$~$ pues, inmediatamente despu\'es de la formaci\'on de esas
primeras estrellas, el enfriamiento at\'omico en
halos de $\sim 10^8 \msun$ empieza a ser activo,
dando lugar a una nueva generaci\'on de galaxias que debi\'o
reionizar el universo.

  Una vez el medio intergal\'actico es reionizado, el enfriamiento
es suprimido por la fotoionizaci\'on, que calienta el gas y disminuye
la abundancia de \'atomos que pueden ser excitados colisionalmente.
La l\'\i nea s\'olida delgada a bajo CR nos da la temperatura en la
cual el tiempo de enfriamiento es menor que el tiempo de Hubble para
el gas ionizado (hemos adoptado aqu\'\i$~$ el modelo de Haardt \& Madau
1996 para el fondo de radiaci\'on ionizante). Las galaxias pueden
formarse solamente por encima de esta l\'\i nea una vez la
reionizaci\'on se ha completado. En realidad, la fotoionizaci\'on tiene
tambi\'en otro efecto: el calentamiento del gas durante la reionizaci\'on
puede producirse antes del colapso, y luego la temperatura del gas sube
adiab\'aticamente cuando la densidad aumenta si el enfriamiento
no es importante (recordemos que las l\'\i neas indicando un tiempo
de enfriamiento igual a la edad del universo son para una sobredensidad
de $18\pi^2 \simeq 178$, el valor obtenido en el modelo esf\'erico
en el momento de la virializaci\'on, y que el tiempo de enfriamiento es
mayor a menor densidad). Debido a este efecto, el gas en halos con
$\sigma \lesssim 30 \kms$ no puede en general enfriarse para formar
galaxias (p.e., Thoul \& Weinberg 1996). Otro efecto que puede
disminuir la eficiencia de formaci\'on de galaxias en halos de baja
dispersi\'on de velocidad es el hecho de que la energ\'\i a liberada por
el propio proceso de formaci\'on estelar puede crear un viento
gal\'actico con energ\'\i a suficiente para calentar y expulsar el gas
en el proceso de acreci\'on (p.e., Dekel \& Silk 1986). Ese proceso
parece ser necesario para evitar un exceso de galaxias de baja
luminosidad comparado con las observaciones (White \& Frenk 1991;
Navarro \& Steinmetz 1997).

\section{Observaciones de galaxias a alto CR}

  Grandes avances en cosmolog\'\i a observacional han tenido lugar
recientemente, con el descubrimiento de gran n\'umero de galaxias a
alto CR. Una de las t\'ecnicas de mayor \'exito es la selecci\'on
fotom\'etrica de objetos d\'ebiles para detecci\'on de la brecha de
Lyman (``Lyman break''; Guhathakurta, Tyson, \& Majewski 1990; Steidel
\etal 1996). Todas las galaxias, cuya luz resulta de la
superposici\'on de los espectros de muchas estrellas, muestran en su
espectro una ca\'\i da abrupta del flujo a la longitud de onda del
l\'\i mite de Lyman, como vemos en los modelos de espectros de distintas
edades con ritmo constante de formaci\'on estelar en la Figura 3, que
reproducimos de Bruzual \& Charlot (1993). La brecha se produce en la
atm\'osfera de las estrellas masivas, que producen la mayor parte de
la radiaci\'on ultravioleta. Adem\'as, la presencia de hidr\'ogeno
at\'omico en el medio interestelar de la galaxia resulta generalmente
en la absorci\'on de la mayor parte de los fotones ionizantes,
aumentando la amplitud de la brecha. A alto CR, la brecha de Lyman se
corre hasta longitudes de onda en el visible, lo que permite
seleccionar las galaxias a partir de sus colores. Por ejemplo, un
objeto azul en V-R y muy rojo en B-V debe tener la brecha entre las
bandas B y V, situ\'andolo a $z\sim 4$.

  El n\'umero de galaxias detectadas por este m\'etodo, de magnitudes
en el rango $23 \lesssim B \lesssim 29$, ha permitido la primera
medici\'on de la funci\'on de luminosidad y de la tasa global de
formaci\'on estelar a alto CR, que resulta ser mucho mayor que la
actual (Madau \etal 1996; Steidel \etal 1999).
La cantidad total de estrellas formadas en estas galaxias, en el
intervalo $1\lesssim z \lesssim 5$, puede dar cuenta de la mayor parte
de la poblaci\'on estelar vieja en el universo presente. Sin embargo,
esas estimaciones est\'an sujetas todav\'\i a a varias incertidumbres:
en general, solamente la tasa de formaci\'on de estrellas masivas se
puede deducir de esas observaciones, y existe la posibilidad de que
la funci\'on de masa inicial fuera distinta a alto CR. Por otra parte,
la tasa total de formaci\'on estelar podr\'\i a ser mucho mayor si
galaxias de baja luminosidad que permanecen por debajo de los
l\'\i mites de detecci\'on dominaran la emisi\'on total, o si la
mayor parte de la radiaci\'on ultravioleta es absorbida por polvo
interestelar y reemitida en el infrarojo lejano (p.e., Calzetti 1999
y referencias inclu\'\i das).

  Efectivamente, la radiaci\'on
reemitida por polvo constituye otra forma de detecci\'on de galaxias
a alto CR. Recientemente, la radiaci\'on de fondo en el infrarojo
lejano, debida a la combinaci\'on de todas las galaxias, ha sido
detectada por COBE (Fixsen \etal 1998; Hauser \etal 1998), y fuentes
individuales se han descubierto con el nuevo detector SCUBA (Eales
\etal 1998; Hughes \etal 1998; Smail \etal 1997), sugeriendo que gran
parte de la radiaci\'on estelar a alto CR pudo ser procesada por
polvo.

  El origen de esas galaxias a alto CR puede comprenderse simplemente
en la teor\'\i a MIF a partir de la Figura 4. Las l\'\i neas punteadas
nos indican aqu\'\i$~$ la magnitud aparente de una galaxia (en el sistema
$AB$, donde la magnitud es $AB = -48.6 - 2.5 log_{10}(f_{\nu})$, y el
flujo $f_{\nu}$ se expresa en unidades cgs), formada en un
halo de dispersi\'on de velocidad $\sigma$ en funci\'on de $z$, para
un modelo de m\'axima luminosidad en que todos los bariones contenidos
en el halo forman estrellas en un intervalo de tiempo igual a la
mitad de la edad del universo en el momento del colapso. En otras
palabras, el modelo representa la m\'axima eficiencia de formaci\'on
estelar posible, donde toda la materia forma estrellas en un tiempo del
orden del tiempo din\'amico del halo de materia invisible
(v\'ease Miralda-Escud\'e \& Rees 1998 para m\'as detalles del modelo).
Las galaxias de mayor masa pueden empezar
a formarse a $z\simeq 4$, cuando el gas puede enfriarse seg\'un la
figura 2b. El flujo m\'aximo de esas galaxias corresponde a $AB \simeq
22$. La mayor parte de las galaxias de la brecha de Lyman son algo
m\'as d\'ebiles, como es de esperar cuando el ritmo de conversi\'on de
gas a estrellas es menos eficiente, y cuando una parte de la
radiaci\'on ultravioleta es absorbida por polvo. Vemos tambi\'en que
las galaxias m\'as masivas pueden formarse m\'as f\'acilmente
cuando el gas ha sido ya enriquecido, debido al mayor ritmo de
enfriamiento a alta metalicidad.

  Puesto que las galaxias detectadas a alto CR de mayor luminosidad
est\'an asociadas con las mayores fluctuaciones de densidad a las
mayores escalas de colapso gravitatorio, su correlaci\'on espacial
deber\'\i a ser mucho mayor que la correlaci\'on de la masa,
como se espera de los altos picos de densidad en un campo
Gausiano (Kaiser 1984).
Esa correlaci\'on ha sido detectada, y es en general consistente con
las expectativas en la teor\'\i a MIF (Adelberger \etal 1998;
Giavalisco \etal 1998; Kauffmann, Nusser, \& Steinmetz 1997). Los
estudios de la correlaci\'on de galaxias a alto CR abren un nuevo
abanico enorme de posibilidades para observar la evoluci\'on de
estructura a gran escala, que hemos solamente empezado a investigar.

  A mayor $z$, la absorci\'on del flujo ultravioleta a $\lambda < 1216\,
{\rm \AA}$ debido al bosque Ly$\alpha$ aumenta r\'apidamente, de tal forma
que a $z\gtrsim 5$, la ca\'\i da de flujo en Ly$\alpha$ se convierte en
la caracter\'\i stica m\'as importante para seleccionar objetos a este
CR. As\'\i$~$ pues, la t\'ecnica para encontrar galaxias a $z > 5$ va
a ser muy parecida a la de la brecha de Lyman, sustituyendo \'esta por
el hoyo de Gunn-Peterson (Gunn \& Peterson 1965). Aunque inicialmente
el t\'ermino del hoyo de Gunn-Peterson se utiliz\'o \'unicamente para
referirse a la absorci\'on en Ly$\alpha$ producida por el medio
intergal\'actico at\'omico antes de la reionizaci\'on, el medio
ionizado puede dar lugar tambi\'en a un hoyo cuando la fracci\'on neutra
es suficientemente grande para que todas las partes del medio
intergal\'actico, incluso las de menor densidad en los vac\'\i os,
absorban esencialmente todo el flujo. Debido al aumento de la densidad
media del gas en el universo, y al mayor ritmo de recombinaci\'on a alto
CR, es de esperar que pr\'acticamente todo el flujo a longitudes de onda
menores que la l\'\i nea Ly$\alpha$ sea absorbido a $z\gtrsim 6$,
incluso si la reionizaci\'on ocurri\'o a mayor CR
(Miralda-Escud\'e \etal 1999).

  Finalmente, otra t\'ecnica importante de detecci\'on de galaxias a
alto CR es mediante la l\'\i nea de emisi\'on de Ly$\alpha$ (Thompson
\etal 1995; Thommes \etal 1998; Meisenheimer \etal 1998; Hu, Cowie, \&
McMahon 1998). En regiones
de formaci\'on estelar, la mayor parte de la radiaci\'on ionizante
emitida por estrellas j\'ovenes es generalmente absorbida por
hidr\'ogeno interestelar, y la energ\'\i a se reemite en fotones de
recombinaci\'on, siendo Ly$\alpha$ la l\'\i nea m\'as brillante.
La b\'usqueda de l\'\i neas de emisi\'on de objetos a alto CR tiene la
ventaja de que el fondo de cielo puede reducirse observando solamente
en una banda estrecha de longitudes de onda, especialmente cuando se
seleccionan longitudes de onda de buena transparencia atmosf\'erica
(especialmente importante a $z\gtrsim 5$, cuando la l\'\i nea Ly$\alpha$
se corre hasta el infrarojo).

  Es evidente a partir de las Figuras 2 y 4 que la dificultad en la
detecci\'on de galaxias a CR progresivamente mayor aumenta r\'apidamente
para $z\gtrsim 5$, puesto que el flujo de las galaxias debe disminuir no
s\'olo debido al mayor CR, sino a la baja luminosidad de las primeras
galaxias, debida a su menor masa comparado con las galaxias actuales.
Esa predicci\'on de la teor\'\i a MIF sugiere que en el futuro, las
fuentes detectadas a mayor CR podr\'\i an ser supernovas, las cuales
debieron ocurrir tan pronto como las primeras estrellas se formaron
(Miralda-Escud\'e \& Rees 1997). La proyectada misi\'on NGST podr\'\i a
detectar supernovas hasta $z\sim 10$ (Stockman \& Mather 1997). Una
posibilidad interesante que podr\'\i a permitir acelerar el estudio
observacional de las primeras estrellas es que los estallidos de rayos
gamma, con sus brillantes contrapartidas \'opticas (p.e., Metzger \etal
1997), sean un fen\'omeno
asociado con estrellas masivas, ocurriendo por consiguiente a los
mayores CR donde exist\'\i an estrellas. Las contrapartidas \'opticas
ser\'\i an detectables en observatorios terrestres, corridas al
infrarojo, mientras que en el visual no habr\'\i a contrapartida debido
al hoyo de Gunn-Peterson y absorci\'on por fotoionizaci\'on. En el caso
de los estallidos de rayos gamma, el reto observacional podr\'\i a
consistir en identificar una peque\~na fracci\'on de esos eventos a
mayor CR que cualquier otra fuente conocida, y distinguirlos de
estallidos a menor CR pero con un gran enrojecimiento debido a polvo
interestelar cerca de la fuente, que pueden tener caracter\'\i sticas
fotom\'etricas similares.

\section{Conclusiones}

  La cosmolog\'\i a observacional est\'a entrando en una nueva etapa de
descubrimiento, con nuevas t\'ecnicas para la detecci\'on de galaxias
d\'ebiles que empujan la frontera de alto CR hacia la \'epoca de la
formaci\'on de las primeras galaxias. Al mismo tiempo, la medici\'on
precisa de las fluctuaciones en la radiaci\'on de fondo, la
realizaci\'on de nuevos escrutamientos de CR de galaxias, y la
continuaci\'on de la b\'usqueda de supernovas a alto CR para la
medici\'on de la geometr\'\i a c\'osmica, prometen fijar los
par\'ametros del universo y del modelo cosmol\'ogico. Hemos visto que
la teor\'\i a MIF reproduce con gran \'exito las observaciones acumuladas
hasta el presente sobre estructura a gran escala. Sin embargo, el estado
actual de la teor\'\i a deja muchas preguntas abiertas: cu\'al es la
naturaleza de la materia invisible? Existe realmente una ``energ\'\i a
de vac\'\i o'' que da cuenta de la densidad necesaria para alcanzar la
densidad cr\'\i tica? Cu\'al es la naturaleza de esta energ\'\i a de
vac\'\i o, cu\'al es su ecuaci\'on de estado, y por qu\'e existe?
Qu\'e proceso gener\'o las fluctuaciones de densidad?
Qu\'e determin\'o su amplitud? Son las fluctuaciones primordiales
exactamente adiab\'aticas y Gausianas, y es su espectro perfectamente
invariante de escala, o existen peque\~nas desviaciones de esta
simple hip\'otesis? Esas preguntas, que nos llevan al misterio de la
\'epoca de inflaci\'on y al origen del universo, ocupar\'an
probablemente el centro de inter\'es en el futuro de la cosmolog\'\i a
observacional, cuyo avance permitir\'a tambi\'en descifrar la historia de
la formaci\'on de galaxias, desde el colapso de las primeras estrellas
hasta el presente.


   Quisiera agradecer David Weinberg por proporcionar un c\'odigo para
calcular el espectro de potencias, y Gustavo Bruzual, Stephane Charlot
y Max Tegmark por permitir la reproducci\'on de resultados de sus
trabajos en este art\'\i culo.
Agradezco tambi\'en diversas
conversaciones con Martin Haehnelt, Martin Rees y David Weinberg.

\newpage

\begin{figure}
\centerline{
\hbox{
\epsfxsize=4.4truein
\epsfbox[55 32 525 706]{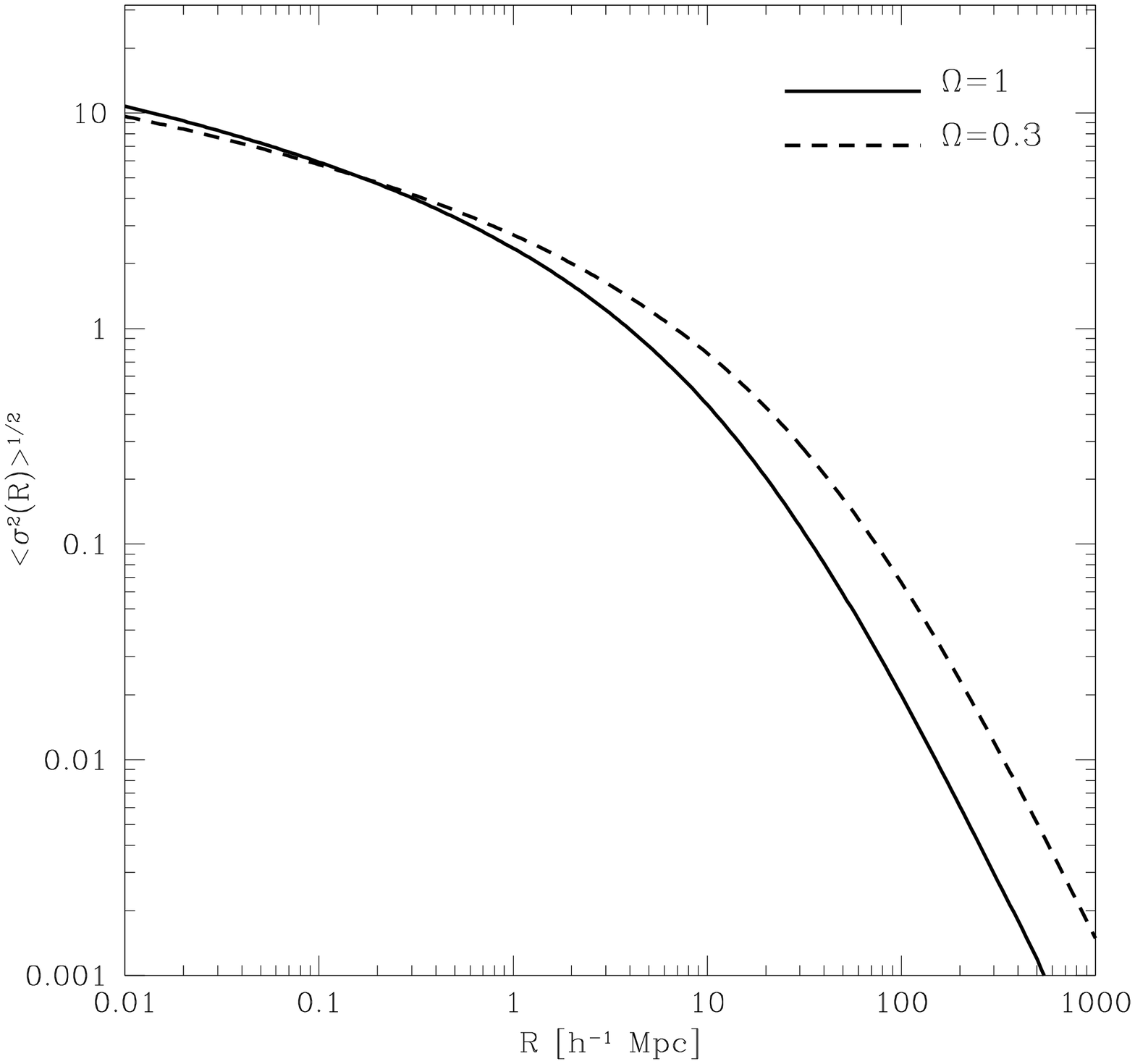}
}
}
\vskip -40pt
\caption{Dispersi\'on en la sobredensidad media dentro de una esfera
de radio $R$ en el campo lineal de densidad para la teor\'\i a MIF con
constante cosmol\'ogica, con dos valores de $\Omega$. }
\end{figure}

\begin{figure}
\centerline{
\hbox{
\epsfxsize=4.4truein
\epsfbox[55 32 525 706]{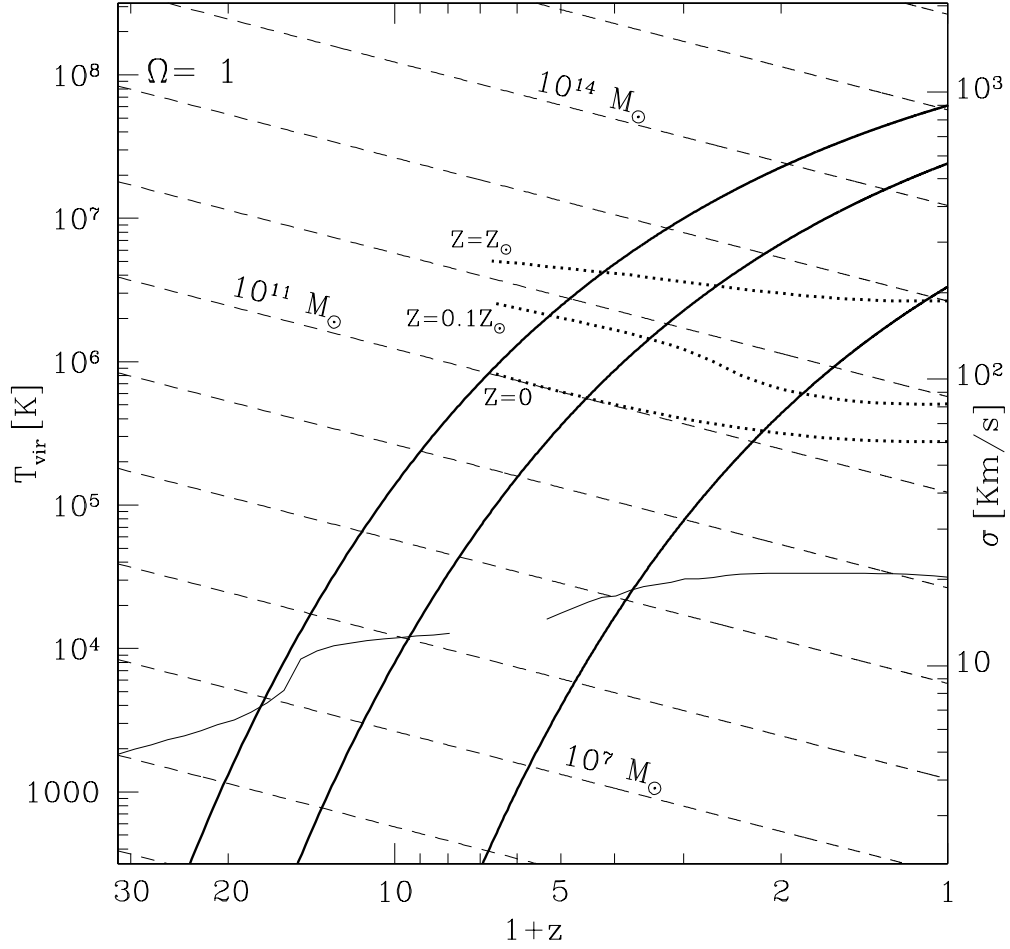}
}
}
\vskip -40pt
\figurenum{2a}
\caption{ Las tres l\'\i neas s\'olidas gruesas dan la dispersi\'on de
velocidades de un halo colapsado a CR $z$, provenientes de fluctuaciones
$(1, 2, 3)\sigma$ en el modelo Press-Schechter. La masa total de los
halos es indicada por la l\'\i nea de rayas. Las l\'\i neas de puntos
indican la m\'axima masa para que el gas pueda enfriarse, para tres
metalicidades distintas, y las l\'\i neas s\'olidas finas indican la
m\'\i nima temperatura para enfriamiento at\'omico y molecular. El
modelo cosmol\'ogico es $\Omega=1$, $\Lambda=0$, $h=0.5$.}
\end{figure}

\begin{figure}
\centerline{
\hbox{
\epsfxsize=4.4truein
\epsfbox[55 32 525 706]{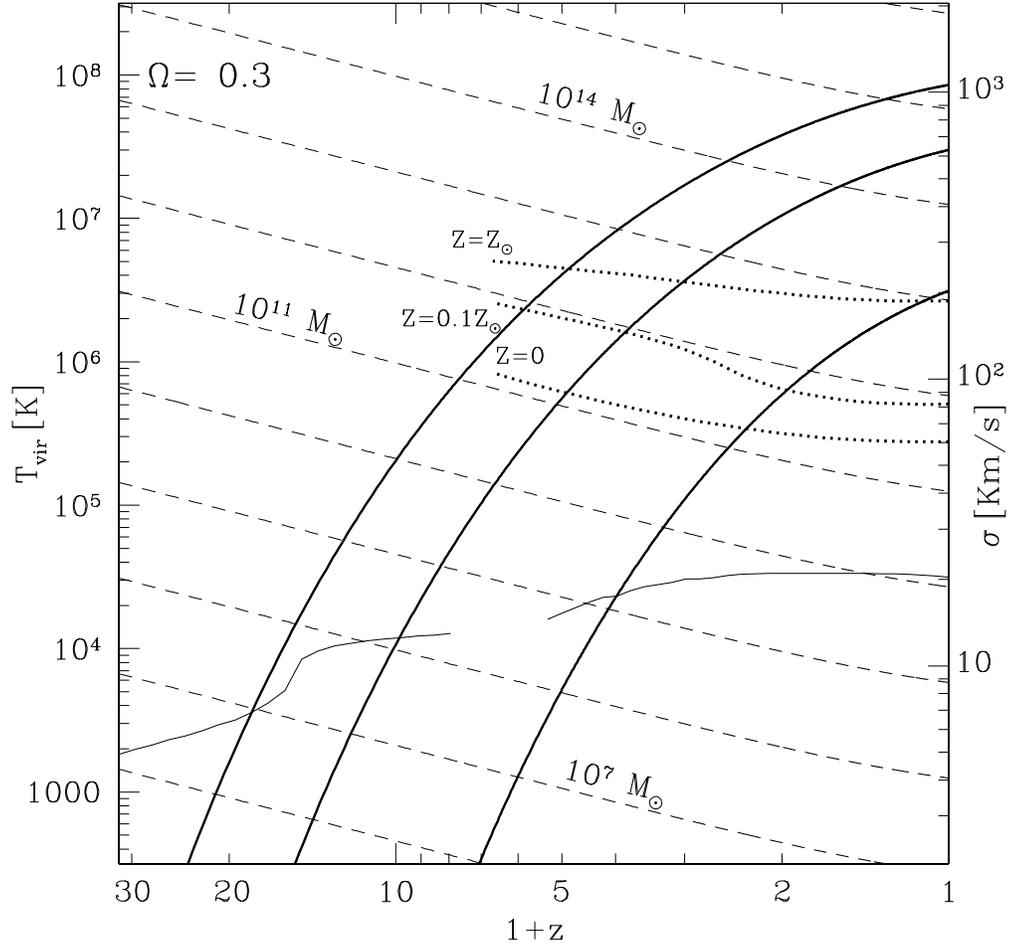}
}
}
\vskip -40pt
\figurenum{2b}
\caption{Igual que en la Figura 2a, para el modelo $\Omega=0.3$,
$\Lambda=0.7$, $h=0.65$.}
\end{figure}

\begin{figure}
\centerline{
\hbox{
\epsfxsize=4.4truein
\epsfbox[55 32 525 706]{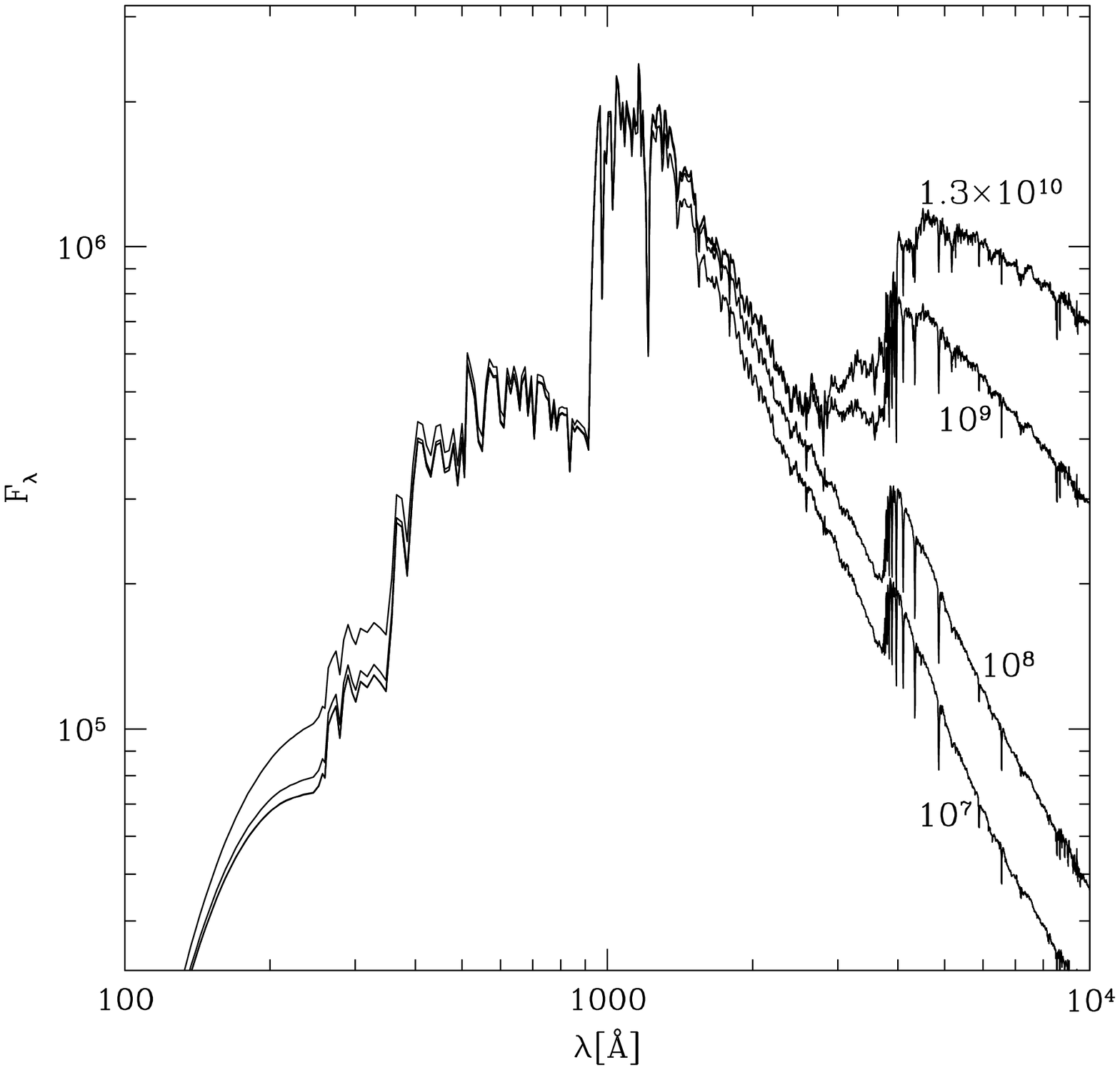}
}
}
\vskip -40pt
\figurenum{3}
\caption{Modelo del espectro de una galaxia con ritmo constante de
formaci\'on estelar, con las edades en a\~nos indicadas. }
\end{figure}

\begin{figure}
\centerline{
\hbox{
\epsfxsize=4.4truein
\epsfbox[55 32 525 706]{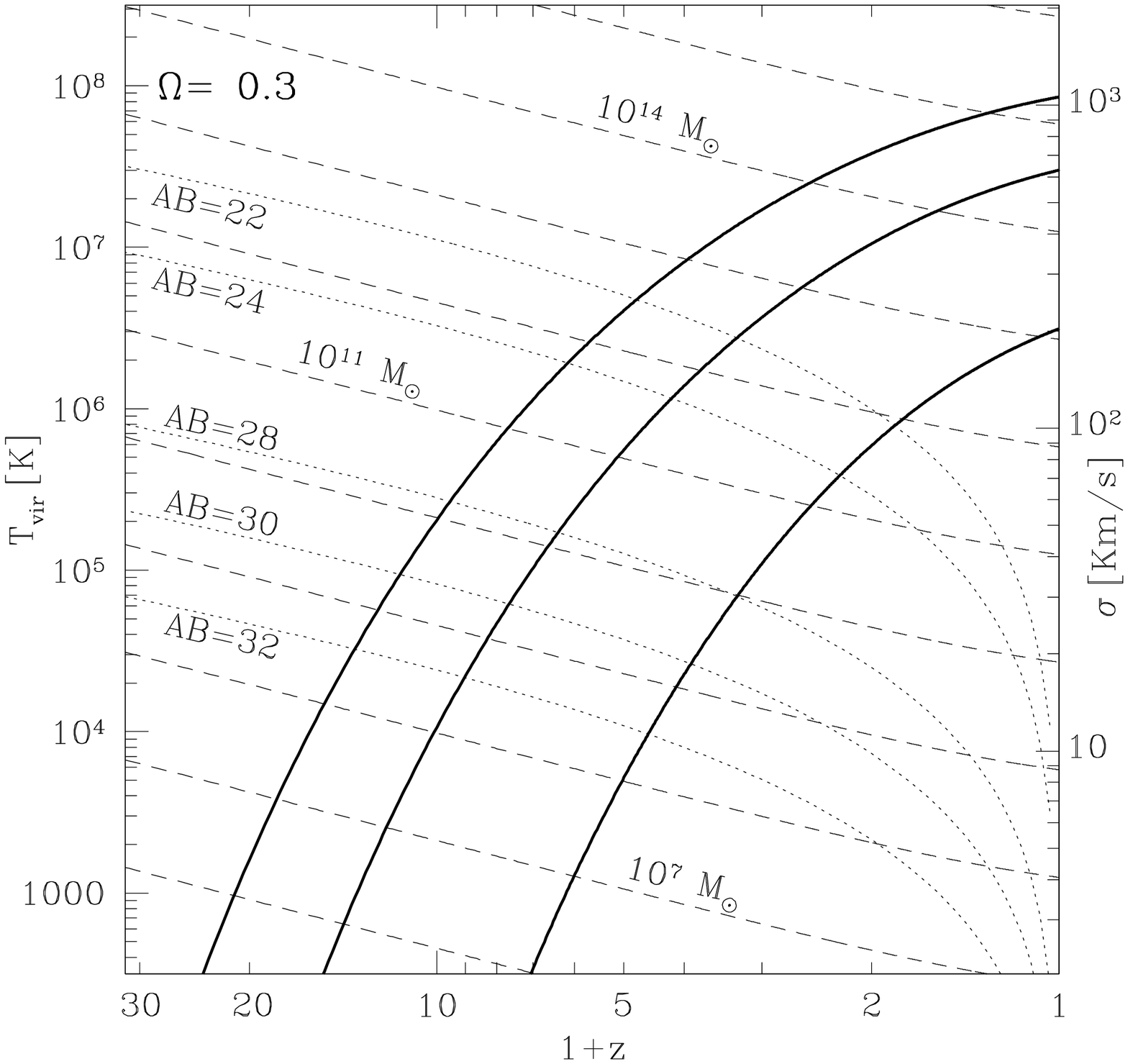}
}
}
\vskip -40pt
\figurenum{4}
\caption{Figura an\'aloga a la figura 2, donde las l\'\i neas de
puntos dan la magnitud AB en una banda ultravioleta suponiendo que
toda la masa bari\'onica en el halo forma estrellas durante un
tiempo igual a la mitad de la edad del universo cuando el halo
colapsa.}
\end{figure}

\end{document}